\documentstyle[12pt]{article}
\begin{document}

\vskip 1truecm
\rightline{Preprint PUP-TH-1531 (1995)}
\rightline{Preprint LANCASTER-TH/9503 (1995)}
\rightline{Bulletin Board hep-ph/9503296}
\bigskip
\centerline{\Large  Bubble wall velocity in a first order
electroweak phase transition}
\bigskip
\centerline{\Large Guy Moore}
\medskip

\centerline{\it Princeton University}
\centerline{\it Physics Department, PO Box 708}
\centerline{\it Princeton, NJ 08544, USA}

\medskip

\centerline{ and }

\smallskip

\centerline{\Large Tomislav Prokopec}

\medskip

\centerline{\it Lancaster  University}
\centerline{\it School of Physics and Chemistry}
\centerline{\it Lancaster LA1 4YB, UK}

\bigskip\bigskip

\centerline{\bf ABSTRACT}

We calculate the velocity and thickness of a bubble wall at the
electroweak phase transition in the Minimal Standard Model.
We model the wall with semiclassical equations of motion and
show that friction arises from the deviation of massive
particle populations from thermal equilibrium.
We treat these  with Boltzmann equations in a
fluid approximation in the background of the wall.
Our analysis improves on the previous work by
 using the two loop
 effective potential, accounting for particle transport, and
determining the wall thickness dynamically.  We
find that the wall is significantly thicker than at phase
equilibrium, and that the velocity is fairly
high, $v_w \simeq 0.7c$, and quite weakly dependent
 on the Higgs mass.

\bigskip

\section{INTRODUCTION}

There has been a growing interest in the idea that
the baryon asymmetry of the Universe may be created
at a first order electroweak phase transition.
However, ingredients needed to construct the whole picture
 are still missing.
The exact nature of the Higgs mechanism is unknown, and
the simplest model, the Minimal Standard Model, apparently
does not contain sufficient CP violation for baryogenesis; we
must consider extensions, such as the two Higgs model.
Even in the Minimal Standard Model we have difficulties computing
the finite temperature effective potential (which is needed to
determine the strength of the phase transition) and the
dynamics of the transition.

Recently there have been advances in
calculating the effective potential;
the two loop contribution has been evaluated \cite{Arnold,Fodor},
 and there has been progress in understanding
nonperturbative effects
\cite{Shapnonpert,Kajantie,Wetterich}.  Both results
support the view that the phase transition is first
order and strong enough to
proceed by bubble nucleation and growth (even when the
Higgs mass
is moderately large, $m_H\sim m_W$).

  To model baryogenesis
accurately one also needs to know the profile
and velocity of an expanding bubble wall.
The wall velocity is friction limited; but determining the
strength of frictive effects involves determining the
non-equilibrium populations of massive
particles in the vicinity of the wall, which
is difficult \cite {Neil,DineLinde,Mac}.  The recent discovery
that the top quark is very heavy \cite{CDF} suggests that top
quarks may be the dominant source of friction, in which case
an approximation which models top quarks with good
accuracy should improve our understanding of the wall motion.
In this letter we re-analyze the bubble wall's velocity and its
shape, working in the Minimal Standard Model and
using the fluid approximation to model the particle populations,
which should treat fermions fairly well.
The technique also allows us to account for transport and to
determine the wall thickness dynamically (taking a specific {\it
Ansatz} for the wall profile).

\section {Equation of Motion for the Higgs vev}

We intend to study the dynamics of infrared condensates in the
Higgs field $\Phi$.  Such condensates should behave
semiclassically
to a good approximation.  From the terms in the Electroweak
Lagrangian containing $\Phi$,

\begin{eqnarray}
{\cal L} =  ({\cal D} _{\mu} \Phi ) ^{\dag} {\cal D} ^{\mu} \Phi
\! -\! V(\Phi ^{\dag} \Phi ) - \sum y ( \Phi ^{\dag}
\overline{{\psi}_{R}}
\psi _{L}\! +\! \Phi \overline{{\psi}_{L}} \psi_{R})
\nonumber
\end{eqnarray}
(where the sum runs over the quarks and leptons $\psi$, and $y$
denotes
the Yukawa  coupling), we find the equations of motion for
$\phi$ (where $\Phi^{\dag} = [0 , \phi / \sqrt{2} ]$) to be
\begin{equation}
 \Box \phi + V' ( \phi) - \frac{g_{w}^{2}}{4} \phi {\rm
Tr}A^{2}  +ig_w A^\mu \partial_\mu \phi
 + \frac{ig}{2} (\partial^\mu A_\mu ) \phi
 + \sum \frac{y}{\sqrt{2}} \bar{\psi} \psi = 0
\label{eq:equation for phi i}
\end{equation}
Here $A$ and $\psi$ are quantum operators.  (For simplicity we
have
set the Weinberg angle $ \tan \theta_W = 0$.) It is reasonable to
take thermal averages of these operators using WKB
wave functions.  This is because the wall will be much thicker
than the thermal length $T^{-1}$, which characterizes the
reciprocal momenta of particles in the plasma.  In this
approximation we get
\begin{equation}
 \Box \phi + V'(\phi ) + \sum {dm^2\over d\phi}
\int \frac{d^{3} k}{ (2 \pi )^{3}\; 2E}  f(k,x) = 0
\label{eq:equation for phi ii}
\end{equation}
where $V$ is the renormalized vacuum
potential, $f$ is the phase space population density (in the
background
of a propagating wall)  and the sum includes all massive
physical degrees of
freedom. Note the condensed notation: $m=y\phi/\sqrt{2}$ for
quarks
and leptons and $m=g_w \phi/2$ for the gauge fields.

To model the population density $f$ we assume a small departure
from the
equilibrium population $f_0$ and write $f = f_0 + \delta f$.
The vacuum contribution $V'(\phi)$ and
the contribution from $f_0$ combine to give the finite
temperature effective potential $V_T ' (\phi)$.  Thus we have
\cite{eqmocomment}

\begin {equation}
\label {eq:equation of motion}
 \Box \phi + V_{T}'(\phi) + \sum { dm^2\over d\phi} \int
\frac{d^{3}p}{(2 \pi )^{3}
\, 2 E}  \delta f(p,x)  =0
\end{equation}
We see that the frictive force arises due to the departure from
thermal
equilibrium $\delta f$.
We will use this equation, an expression for $V_T$,  and
 equations for  $\delta f$
to compute the wall velocity and its shape once it has reached
a planar steady state, $\phi = \phi ( z+v_w t) , \delta f =
\delta f ( z+v_w t)$.


\section {Effective Potential}

The high temperature expansion of the one loop effective
potential is \cite{DineLinde}
\begin {equation}
 V_{T}(\phi) = D (T^{2} - T_{0}^{2} ) \phi ^{2} - E \phi ^{3} T
+\frac{ \lambda_{T}}{4} \phi ^{4}
\end{equation}
with $D = (2 m_{t}^{2} + 2 m_{W}^{2} + m_{Z}^{2})/8v_{0}^2
\simeq .167$, $E= (2m_{W}^{3}
+m_{Z}^{3} )/ (4\pi v_{0}^3) \simeq 0.01$,  $\lambda_T$ the
Higgs self coupling at a scale roughly given by T,
 and $T_0= [ m_H^2
+3(4m_t^4 - 2 m_W^4 - m_Z^4 ) / 8 \pi^2 v_0^2 ] /4D$.
 Recently the authors of \cite {Arnold} and \cite{Fodor} have
computed
the two loop expression.  The most important changes are that
$E$ becomes $(4m_W^3 + 2m_Z^3 + 3( 1 + \sqrt{3}) \lambda^{3/2})
 / (12 \pi v_0^3)$ and a qualitatively new
term $-B \phi^2 T^2 \log(\phi / T)$ appears.  The result of
\cite{Fodor} is  $ B \simeq (1.4g_{w}^{4} -.48g_{w}^{3}
\sqrt{\lambda}+5.1g_{w}^{2} \lambda - 6.7 \lambda^{2})/(16 \pi
^{2}) $.
It may also be important to include nonperturbative effects.
Shaposhnikov proposes a term $-(A_{f} g_{w}^{6} T^4/12)\; {\rm
Pit}(\phi)$
\cite{Shapnonpert}.  The function Pit describes
the contribution of a gauge condensate and it is roughly
constant near $\phi = 0$
 and falls exponentially as $\,\exp -\phi/g^2 T $ for large
$\phi$. We  add such a term to parametrize our ignorance about
the free energy of the symmetric phase;
we use $-(A_{f} g_{w}^{6} T^4 /12)\; {\rm sech}(32
\phi / T)$ because it is simple -- the exact form
of Pit will have little effect on our calculations
 -- and we take $\lambda_T$ and $A_{f}$ as unknowns.

What should we use for $T$?  Most of space
is converted to the broken phase by bubbles which nucleate when
the critical bubble free energy reaches $S\simeq 100 T$
\cite{AndersonHall}.  We have computed
$S$ using our form for $V_T$ and the techniques used
in \cite{Neil,DineLinde,Mac} and find, for
instance, that for $\lambda_T=0.03$, $A_f = 0$, and $S=100 T$
that
$T = 1.00006 T_0$.  (Note that because of the $B$ term,
$T_0$ is no longer the spinodal temperature.)

\section{Fluid Equations}

Next we must determine $\delta f$, the deviation from equilibrium
in the presence of the moving wall.  Our starting point is the
Boltzmann equation,
\begin{equation}
\partial_t f +{p_z\over E}\partial_z f+{\dot p_z}\partial_{p_z}
f =-C[f]
\label{eq:boltzmann}
\end{equation}
where $C[f]$ represents the scattering integral,
$E=(p^2+m^2)^{1/2}$ is
the particle energy, $v_z=p_z/E$ is the velocity (with the broken
phase at positive $z$), and $\dot p_z=-\partial_z E$ is the
force on
the
particle.  In a complete description  each particle species in
the
plasma would be described with a Boltzmann equation.
We allow ourselves the
approximation that all species but top quarks and perhaps $W$
bosons are in equilibrium
and we neglect the slight change in the background temperature
across the wall.  The former approximation is
reasonable as the induced
deviation from equilibrium goes as $m^2$ and top quarks and
$W$ bosons are the heaviest particles.  We can correct for the
latter approximation by using the work of
 Enqvist {\it et al.\/} \cite{IgnatiusKKL} and Heckler
\cite{Heckler}, who have used hydrodynamic conservation laws
to relate wall velocity to entropy production;
for simplicity we will not do so here.
We also use a single $f$ for
tops and antitops of both helicity.  In the Minimal Standard
Model this is reasonable as there is almost no CP violation, and
the difference in transport properties arises only at the
subleading
level of weak scatterings.
In two doublet models we might need to be more careful.

The Boltzmann equations are nonlinear partial
integro-differential
relations and  as such are analytically intractable.  To solve
them we  make the fluid {\it Ansatz}, assuming $f$ to be of the
form

\begin{equation}
f={1\over 1+ {\rm exp}{ {E-  E \delta T/T - p_z v
 -\mu\over T}}}
\label{eq:distributionfunction}
\end{equation}
where we have written explicitly three types of perturbations:
chemical
potential $\mu$, temperature $\delta T$  and velocity
 $v$.  This {\it Ansatz} is a truncation of an expansion
in powers of momentum; it gives a reasonable description
of the populations of thermal energy particles when the
background
varies slowly on the scale of the diffusion length.
For top quarks this
should be sufficient as the diffusion length is short and the
influence of infrared particles is phase space suppressed.  For
$W$ bosons, Bose statistics give large infrared particle
 populations, and the fluid approximation is unreliable
unless $W$ bosons thermalize on time
scales short compared to their annihilation rate.  We consider
the fluid equations for $W$ bosons as a guide for their
importance
and concentrate on top quarks.
We will furthermore work
to linear order in perturbations which are of order $(m / \pi
T)^2$,
and therefore naturally small.

With a three parameter {\it Ansatz\/}
(\ref{eq:distributionfunction}) we cannot ask that the
full Boltzmann equations be satisfied, but only impose that
three
moments be satisfied, namely the integrals over
$\int {d\!\!\!^-}^3 p$, $\int {d\!\!\!^-}^3 p \: E$, and $\int
{d\!\!\!^-}^3 p \: p_z$.
Working in the fluid frame and using $\partial_t f(z+v_w t)=v_w
f'$
 we obtain the following equations \cite{JPT}:
\begin{eqnarray}
 a v_w {\mu ' \over T}+ v_w{ \delta T' \over T} + {1\over
3}
v' +F_1 & = & -\Gamma_{\mu1} \frac{\mu}{T}
-\Gamma_{T1} \frac{\delta T}{T}
\label{eq:fluidequationi} \\
b v_w {\mu '\over T} + v_w {\delta T' \over T} +
{1\over 3} v' +F_2 & = &
-\Gamma_{\mu 2}  \frac{\mu}{T}
-\Gamma_{T2} {\delta T\over T}
\label{eq:fluidequationii}  \\
b {\mu' \over T}\; +\; {\delta T' \over T} \: + v_w v'
+ \: 0 \: & = &
-\Gamma_v  v
\label{eq:fluidequationiii}
\end{eqnarray}
where $a=2\zeta_2/9\zeta_3$,  $b=3
n_0/4\rho_0=3\zeta_3/14\zeta_4$,
$\zeta$ is the Riemann $\zeta$-function: $\zeta_2=\pi^2/6$,
$\zeta_3\approx 1.202$, $\zeta_4=\pi^4/90$, $n_0=3\zeta_3
T^3/4\pi^2$,
$\rho_0=21\zeta_4 T^4/8\pi^2$.

The force terms are:
\begin{equation}
F_1
   =-{v_w \ln 2\over 9\zeta_3} { (m^2)' \over T^2} \,,\qquad
F_2
   =-{v_w \zeta_2\over 42\zeta_4}{ (m^2)' \over T^2}
\label{eq:force}
\end{equation}
They drive the plasma out of equilibrium, while  scatterings
restore it.
Scatterings keep the plasma near equilibrium when the wall
passage
time $L/v_w \gg \Gamma^{-1}$.

We have computed the coefficients on the {\it r.h.s.} of
(\ref{eq:fluidequationi}) - (\ref{eq:fluidequationiii})
including all diagrams which contribute to order $\alpha_{s}^{2}
\log{\alpha_s}$. We found  \cite{tobe}
$\Gamma_{\mu 1}\approx T/26$,
$\Gamma_{T 1}\approx T/14$,
$\Gamma_{\mu 2}\approx T/60$,
$\Gamma_{T 2}\approx T/16$,
$\Gamma_{v 1}\approx T/13$,
where we used  $\alpha_s\equiv\alpha_s(m_Z)\simeq 0.12$.  For
$W$ bosons the values are about half as large.

The force terms are proportional to $m^2$, so in
general $\delta f \propto m^2$.  Note that the friction term in
Eq. (\ref{eq:equation of motion}) is proportional to  $m^2
\delta f \propto m^4$.
For bosons the coefficients in Eqs.
(\ref{eq:fluidequationi}) -- (\ref{eq:force}) differ, and in
particular
$F_1 \propto m^2 \log(2/m)$. The integral in the equation of
motion involving $\mu$ also contains a log enhancement; but
because $ m_{t}^{4} / m_{w}^4$ is very
large, $W$ bosons still produce less friction than top quarks.

\section{Computing Velocity and Profile}

With our {\it Ansatz} for $f$ we can
rewrite
Eq. (\ref{eq:equation of motion}) for top quarks (with
$\sum\rightarrow 12$) as
\begin {equation}
-(1-v_{w}^{2}) \phi '' + V_{T}'(\phi) +
12\frac{dm^2}{d\phi} T \left [ {\mu \ln 2 + \zeta_2 \delta T
\over 4\pi^2} \right ]\! =\! 0
\label {eq:eomforphii}
\end{equation}
This equation and the fluid equations form a system of
nonlinear differential equations for
the wall profile and velocity.  We will attempt to solve them in
\cite{tobe}, but here we will content ourselves with an
{\it Ansatz} for $\phi$,
\begin{equation}
\phi={\phi_0\over 2}\left (1+\tanh {z + v_w t\over L}  \right )
\label{eq:higgsvev}
\end{equation}
where $\phi_0$ is the value of $\phi$ in the asymmetric phase
and $v_w$ the wall velocity
and $L$ the wall thickness in the plasma frame are treated as
undetermined parameters. This {\it Ansatz} is chosen because the
static equilibrium wall shape in the one loop approximation is of
this form.

Again, because we have restricted the form of $\phi$ we cannot
ask that the full equations of motion be satisfied; we can only
enforce two moments.  The natural choices are the space integral
of the equation of motion times $\partial \phi / \partial
v_w$ and $\partial \phi / \partial L$.
Note that
$\partial\phi/\partial v_w =t\phi '$ and
$\partial\phi/\partial L =-(x/L)\phi '$,
so an equivalent set of conditions is
\begin{equation}
\int [{\rm Eq.}\,\, (\ref{eq:eomforphii})] \phi' dz=0
\,,\quad
\int [{\rm Eq.}\,\, (\ref{eq:eomforphii})] \frac{x}{L} \phi '
dz=0
\label{eq:constraints}
\end{equation}
These equations have a simple physical interpretation.
The first equation is the
total pressure on the wall in its rest frame \cite{Neil};
if it were non-zero the wall would accelerate, changing $v_w$.
The second equation is the asymmetry in the pressure between the
front and back edges of the wall; if it were nonzero the wall
would
be compressed or stretched, changing $L$.

The integrals for the first two terms in (\ref{eq:eomforphii})
are
\begin{eqnarray}
 \int  ( \Box \phi + V_{T}'(\phi)) \phi' & = & V_T (\phi_0) - V_T
(0)
\equiv - \Delta V_T \label{eq:constraint one}\\
 \int  [ \Box \phi + V_{T}'(\phi) ] \frac{z}{L} \phi' & = &
\frac{(1-v^2_w) \phi^2_0}{6L^2} - \frac{1}{2}[ \Delta V_T + \Xi
]
 \label{eq:constraint two}\\
 \Xi  \equiv   B\phi_0^2
(\zeta_2 - 1) T^2 & + & \frac{E\phi_0^3 T}{2}  - \frac{5\lambda_T
\phi_0^4}{24}
   + \frac{A_f g_w^6 T^4}{12}\left (2.82 + {1\over 2} \ln
{\phi_0\over
T}\right ) \label{eq:constraint three}
\end{eqnarray}
Note that the $\Box \phi$ term acts to stretch the wall (increase
$L$) while $V_T$ acts to accelerate and
compress the wall.
The coefficient 2.82 in the last term is the only place where
our choice for the function Pit enters our computation.

We will first get a rough estimate of the wall velocity and
thickness by solving Eqs. (\ref{eq:fluidequationi}) --
(\ref{eq:fluidequationiii}) and  (\ref{eq:constraints})  ignoring
transport, by which we mean we will ignore the derivative terms
on the l.h.s.\ of the fluid equations.
Transport reduces friction because
particles tend to flow off the wall, where they contribute less
to equations (\ref{eq:constraints}).
We will also ignore $\delta T$, which turns out to be a
good approximation.
The expression for $\mu$ now becomes rather
simple:
\begin{equation}
\mu = v_w {\ln 2\over 9\zeta_3} \, y_t^2 \, {\phi\phi'\over
\Gamma_{\mu 1}T}
\label{eq:mu no transport}
\end{equation}
Note that $\mu$ does not depend on other decay rates apart from
$\Gamma_{\mu 1}$.

Using $\int (\phi\phi')^2=\phi_0^4/10 L$,
$\int x (\phi\phi')^2=\phi_0^4/24$  and
(\ref{eq:constraint one}) -- (\ref{eq:constraint three}) one can
solve
for $L_w=L/ (1-v_w^2)^{1/2}$ and $v_w$:
\begin{eqnarray}
{1\over L_w^2} = \frac{1}{\phi_0^2} \left( {\Delta V_T \over 2}
+ 3 \:\Xi \right ) \qquad
\gamma_w v_w  =  {15 \pi^2 \zeta_3 \over 2 \ln^2 2}\;
 \Gamma_{\mu 1} L_w { \Delta V_T \over m_t^4}
\label{eq:L no transport}
\end{eqnarray}
For $m_t=174$GeV, $\lambda_T=0.03$ and $A_f = 0$ these give
 $L_w \simeq 29 T^{-1}$
and $\gamma_w v_w \simeq 1.1$, a mildly relativistic and fairly
thick wall.  Note that $L \simeq 20T^{-1}$ is much thicker than
the top quark diffusion constant $D \simeq 4 T^{-1}$, so the
fluid approximation is in good shape.

Now we solve the problem including transport.
First we must find the contributions to (\ref{eq:constraints})
 involving $\mu$ and $\delta T$.
This is most easily accomplished by Fourier analysis.
Let us write Eqs. (\ref{eq:fluidequationi}) --
(\ref{eq:fluidequationiii}) in a matrix notation,
\begin{equation}
A\delta ' + \Gamma \delta = F \, \phi \phi'
\label{matfluid}
\end{equation}
where $\delta$ is a column vector of $\mu,\, \delta T, \, v$,
$A$ is the matrix of coefficients for the derivative terms,
$\Gamma$ is a matrix of the decay constants, and $F$ is a column
vector of the coefficients for the force terms.  Note that $A$
is velocity dependent, and $F$ is linear in velocity.  In Fourier
space (\ref{matfluid}) becomes
\begin{equation}
ik \delta + A^{-1}\Gamma \delta = A^{-1}F \,
 \widetilde{\phi\phi'}
\end{equation}
which may be solved by eigenvalue methods.  Denoting the
eigenvalues and eigenvectors of $A^{-1}\Gamma$ as $\lambda_i$
and $\xi_i$ and expanding $A^{-1}F = \alpha_i \xi_i$, we find
\begin{eqnarray}
\delta & = & \sum_{i} \frac{\alpha_i}{\lambda_i + ik} \xi_i \,
\widetilde{\phi \phi'} \\
\widetilde{\phi \phi '} & = & (\phi_0^2 /2)
\left( 1-\frac{ikL}{2} \right)\left( \frac{kL\pi}{2} \right)
{\rm csch}\left( \frac{kL\pi}{2} \right)
\end{eqnarray}
This gives an explicit expression for $\tilde{\mu}$ and
$\tilde{\delta T}$.

Finally, we can convert the relevant integrals
of the friction term
\begin{eqnarray}
 \frac{3 y_t^2 T}{\pi^2} \int [ \mu(x) \ln 2 + \zeta_2 \delta
T(x)]
 \phi \phi' (x) dx \nonumber \\
 \frac{3 y_t^2 T}{\pi^2} \int [ \mu(x) \ln 2 + \zeta_2 \delta
T(x)]  \frac{x}{L}
 \phi \phi' (x) dx
\label{friction}
\end{eqnarray}
into k-space integrals using the relations
$\int f_1(x) f_2(x) dx =$ $\int \tilde{f_1}(-k) \tilde{f_2}(k)
dk/2\pi$
and $xf(x) \Rightarrow i d\tilde{ f}(k)/ dk$.  This yields
integrals
of form
\begin{equation}
\int \frac{1}{\lambda + ik} \left(1+\frac{k^2L^2}{4} \right)
\left( \frac{kL\pi}{2} {\rm csch} \frac{kL\pi}{2} \right)^2
\frac{dk}{2\pi}
\end{equation}
which may be converted to a rapidly converging infinite sum by
residue integration, or performed numerically.

This completes the evaluation of all terms in Eqs.
(\ref{eq:constraints}).
These equations each define a curve in
the space of $v_w$ and $L$.  The intersection of these curves is
a self consistent solution for the wall shape and velocity within
the {\it Ans\"{a}tze} and approximations we have made.

\section{Results}

We have solved these simultaneous conditions for some
representative
values of $\lambda _T$ and $A_f$.  We find that the friction from
$W$ bosons, calculated in the fluid approximation, is about
half that from top quarks. Though our techniques are different
than those of \cite{Mac}, we get a similar numerical value for
the
friction from $W$s. We have included them in our analysis.
Using
the temperature where the critical bubble action $S=100T$,
we find \[
\begin{array}{llllll}
\lambda_T   &  0.02 \; \; & 0.03 \;\;  &   0.05\; \;& 0.05\; \;&
0.03
\\ A_f    &    0     &    0     &    0    &  0.1  &  0.1
\\ \phi_0/T& 1.06    &   0.78   &   0.57  &  0.78 &  1.03
\\ v_w    &  .84     &   .68    &    .66  &  .96  & {\rm no\:
solution}
\\ T*L     &  43     &   29     &    23   &  9.4  & {\rm no\:
solution}
\\ v_{\rm no \:str}&.33  &   .39    &  .48    &  .68  &  0.54
\\ T*L_{\rm no \: str}&15.6&   16.5   &  16.5   &  7.3  &  9.6
\end{array} \]

The last two columns are the velocity and thickness of the wall
when we treat $v_w$ as a free parameter but fix $L$ to the
value derived from (\ref{eq:constraint two}) without the
contribution from friction.  This value of $L$ approximately
equals the thickness at phase equilibrium.
We see that the wall
is significantly deformed by the frictive effects, and that this
increases its velocity.
Of course, if the deformation is large then we have little
reason to believe that our wall shape {\it Ansatz} is
accurate -- one should model the shape more carefully than
we have done here.
The conclusion that the wall is fast and thick should be
reliable, however.

When we include a sizeable value of $A_f$, the parameter
describing
nonperturbative symmetric phase effects, we find no solution.
The two equations (\ref{eq:constraints}) turn out to be
incompatible;
the wall runs away, but maintains finite plasma frame thickness.
This result probably comes from neglecting friction from the
gauge condensate responsible for $A_f$, which would compress the
wall and prevent runaway.  To remedy this shortcoming we need a
model
for the nonequilibrium dynamics of nonperturbative infrared
condensates.

The situation in two doublet models may be quite different from
what we have found here.  In these theories there are several new
massive (Higgs) bosons.  The ones which do
not couple to the top quark have quite long half-lives and
sizable diffusion constants, and may be a major source of
friction.

\centerline{\bf Acknowledgements}

We thank the Sir Isaac Newton Institute for Mathematical
Sciences,
Cambridge, England, for hospitality during the early part of this
work.  GM  acknowledges support from a National Science
Foundation Graduate Fellowship, and  TP acknowledges funding
from
PPARC. We thank Andrew Heckler, Michael Joyce, and Neil Turok
for
useful discussions.


\begin{thebibliography}{99}
\bibitem{eqmocomment} Ref. \cite{Neil} contains an integral form
 of this expression; it is derived in a careful field theoretic
 way in \cite{Mac}.
\end{thebibliography}
\end{document}